# 基于混合数据驱动算法的SCR氮氧化物排放量动态预测模型

唐振浩[1]，王世魁[1]，曹生现[1]，李扬[2]，沈涛[3]

（1．东北电力大学自动化工程学院，吉林省 吉林市 132012；2．东北电力大学电气工程学院，吉林省 吉林市 132012；3．哈尔滨锅炉厂有限责任公司，黑龙江省 哈尔滨市 150040）

## Dynamic Prediction Model for NOx Emission of SCR System Based on Hybrid Data-driven Algorithms

TANG Zhenhao[1]，WANG Shikui[1]，CAO Shengxian[1]，Li Yang[2], SHEN Tao[3]

(1. School of Automation Engineering, Northeast Electric Power University, Jilin 132012, Jilin Province, China; 2. School of Electrical Engineering, Northeast Electric Power University, Jilin 132012, Jilin Province, China; 3. Harbin Boiler Company Limited, Harbin 150040, Heilongjiang Province, China)

**ABSTRACT:** Aiming at the problem that delay time is difficult to determine and prediction accuracy is low in building prediction model of SCR system, a dynamic modeling scheme based on a hybrid of multiple data-driven algorithms was proposed. First, processed abnormal values and normalized the data. To improve the relevance of the input data, used MIC to estimate delay time and reconstructed production data. Then used combined feature selection method to determine input variables. To further mine data information, VMD was used to decompose input time series. Finally, established NOx emission prediction model combining ELM and EC model. Experimental results based on actual historical operating data show that the MAPE of predicted results is 2.61%. Model sensitivity analysis shows that besides the amount of ammonia injection, the inlet oxygen concentration and the flue gas temperature have a significant impact on NOx emission, which should be considered in SCR process control and optimization.

**摘要：**针对火电厂选择性催化还原(selective catalytic reduction，SCR)系统建模中存在的时延难确定、模型精度不高等问题，提出了一种基于多数据驱动算法混合的动态建模方案。首先，处理原始生产数据异常值并进行归一化。为提高输入数据相关性，使用最大信息系数(maximal information coefficient，MIC)估算各变量的延迟时间，对数据重构。然后采用组合特征选择方法确定输入变量。为了进一步挖掘数据信息，使用变分模态分解 (variational mode decomposition，VMD)对输入时间序列进行分解。最后，建立结合极限学习机(extreme learning machine，ELM)和误差修正(error correction，EC)模型的 NOx 排放量预测模型。基于实际历史运行数据的实验结果表明所建立模型预测结果的 MAPE 为 2.61%。模型敏感性分析表明，除喷氨量外，入口氧气浓度及烟气温度对 NOx 排放量存在显著影响，在 SCR 过程优化控制中应重点考虑。

## 0 引言

氮氧化物(nitrogen oxides，NOx)排放造成严重环境污染且危害人类健康。选择性催化还原(selective catalytic reduction，SCR)系统由于安装简便、脱硝效率高等特点被广泛应于电站锅炉 NOx 排放物的后处理过程中[1]。为了实现对 SCR 系统的高效控制，需要建立 SCR 出口 NOx 排放量的准确预测模型[2-4]。但是，SCR 系统具有非线性、大滞后性和强扰动的特点[5-6]，使得 SCR 系统建模成为一个具有挑战性的问题。

根据建模原理的不同，SCR 系统建模的方法主要分为基于燃烧机理建模和基于实际生产数据建模两种。基于燃烧机理的模型通过根据锅炉系

基金项目：国家自然科学基金(61503072)，吉林省科技发展计划项目(20190201095JC, 20200401085GX)。

The National Natural Science Foundation of China (Grant No.: 61503072) and Jilin Science and Technology Project (Grant No.: 20190201095JC, 20200401085GX).

统内的化学反应和能量变换构建相应微分方程模型。姚楚等人[7]根据 SCR 系统内的化学反应原理构建出口 NOx 排放量的预测模型。Devarakonda 等人[8]基于反应机理建立了 SCR 系统的动力学模型，并分析多种因素对 SCR 系统脱硝效率的影响。但是，由于燃烧过程的复杂性和多变性，机理建模需要设置多种实际生产中难以满足的假设条件，并且模型参数多，计算复杂，难以应用到实际生产中进行在线快速预测[9]。而基于实际生产数据建模方法利用电厂历史运行数据拟合 SCR 出口 NOx 浓度和其余变量之间的非线性关系，更加简单易用，其主要包括统计学回归模型和人工神经网络两类。刘吉臻等人[10]提出了核偏最小二乘(kernel partial least squares，KPLS)模型结合多核学习、模型更新的建模方法，并用于 SCR 烟气脱硝系统建模。朱钰森等人[11]针对循环流化床锅炉燃烧过程中 NOx 生成量的建模问题，提出了基于最小二乘支持向量机(least square support vector machine，LSSVM)加权连接的多模型建模方案。基于人工神经网络的预测模型也被成功应用到电站生产中。Lv 等人[12]使用遗传算法选择典型运行数据，并基于人工神经网络(artificial neural network，ANN)预测 SCR 出口 NOx 浓度。Xie 等人[9]利用长短期记忆(long short-term memory，LSTM)神经网络处理时间序列数据的能力，预测未来时刻 SCR 系统出口 NOx 浓度。温鑫等人[13]利用深度双向 LSTM(deep bidirectional LSTM，DBLSTM)学习建模数据集的深层时间特征，建立 SCR 系统 NOx 排放量的预测模型。这些算法的成功应用证明了基于实际生产数据构建 NOx 浓度模型的可行性。

由于 SCR 系统脱硝过程表现出大延迟的特点，分布式控制系统(distributed control system，DCS)当前时刻相关参数的记录值并不能准确对应 SCR 系统出口 NOx 浓度，因此系统内相关参数的延迟时间成为影响建模精度的重要因素。Yang 等人[14]基于互信息计算 SCR 入口 NOx 浓度和锅炉燃烧过程中相关参数之间的延迟时间。Lv 等人[15]在构建电站湿法烟气脱硫系统中氧化还原电位(oxidation reduction potential，ORP)的动态预测模型时，考虑将 ORP 及输入参数的延迟时间作为优化变量，以模型预测误差为粒子群优化算法的适应度函数，求解相关参数的延迟时间。但是上述确定延迟时间的方法会导致输入变量维度的增加及输入变量之间耦合性的增强，考虑到相关参数历史时刻序列和 SCR 出口 NOx 浓度之间的非线性相关程度，所提出算法基于最大信息系数[16] (maximal information coefficient，MIC)计算非线性相关程度最大的历史时刻，从而计算相关参数的延迟时间。同时，由于 SCR 脱硝反应受到机组负荷、温度、催化剂等众多因素影响，因此 SCR 脱硝系统中的历史运行数据具有非平稳性、随机性等特点，导致建模数据中深层时频信息不明显。变分模态分解[17] (variational mode decomposition，VMD)作为自适应的信号分解方法，使用迭代求解变分模型以确定各模态分量的频率中心值和带宽值，从而完成信号频域空间剖分及子分量的分离，突显数据的局部特征并挖掘数据深层时频信息[17-18]。VMD 算法已经在电力负荷预测[19-20]、风电故障分析[21]等问题中进行了有效的数据分解。因此，本文所提出算法基于 VMD 算法处理建模数据，挖掘其中的深层时频信息。另外，研究设计误差修正策略，通过挖掘预测值和测量值之间误差序列的规律性，建立误差预测模型，对初始的预测值进行误差补偿，进一步提升预测模型的精度。

综上，为了建立 SCR 系统的动态模型，提出了一种基于混合数据驱动算法的动态预测模型。首先，使用最大信息系数分析数据之间非线性相关性，确定相关参数的延迟时间，对实际生产数据进行重构；然后通过组合特征选择策略确定模型输入变量；进一步，对加氨量进行 VMD 分解，深入挖掘数据中的深层时频信息。最后，建立基于极限学习机预测模型，并对模型进行误差修正，实现对 SCR 出口 NOx 浓度的高精度预测。

## 1　SCR 脱硝系统

国内某电厂 1000MW 超超临界燃煤机组的脱硝装置的内部结构如图 1 所示。脱硝装置的核心设备是 SCR 反应器，其余设备包括吹灰器、催化剂、烟气系统等。在燃烧过程中产生的烟气水平进入反应器顶部，然后垂直向下穿过反应器。首先，来自氨站的氨气和由稀释风机提供的空气进行混合并进入到 SCR 反应器内。在催化剂的作用下和入口烟气发生选择性催化反应，该化学化学反应将有害的氮氧化物转化成水和氮气[22]，从

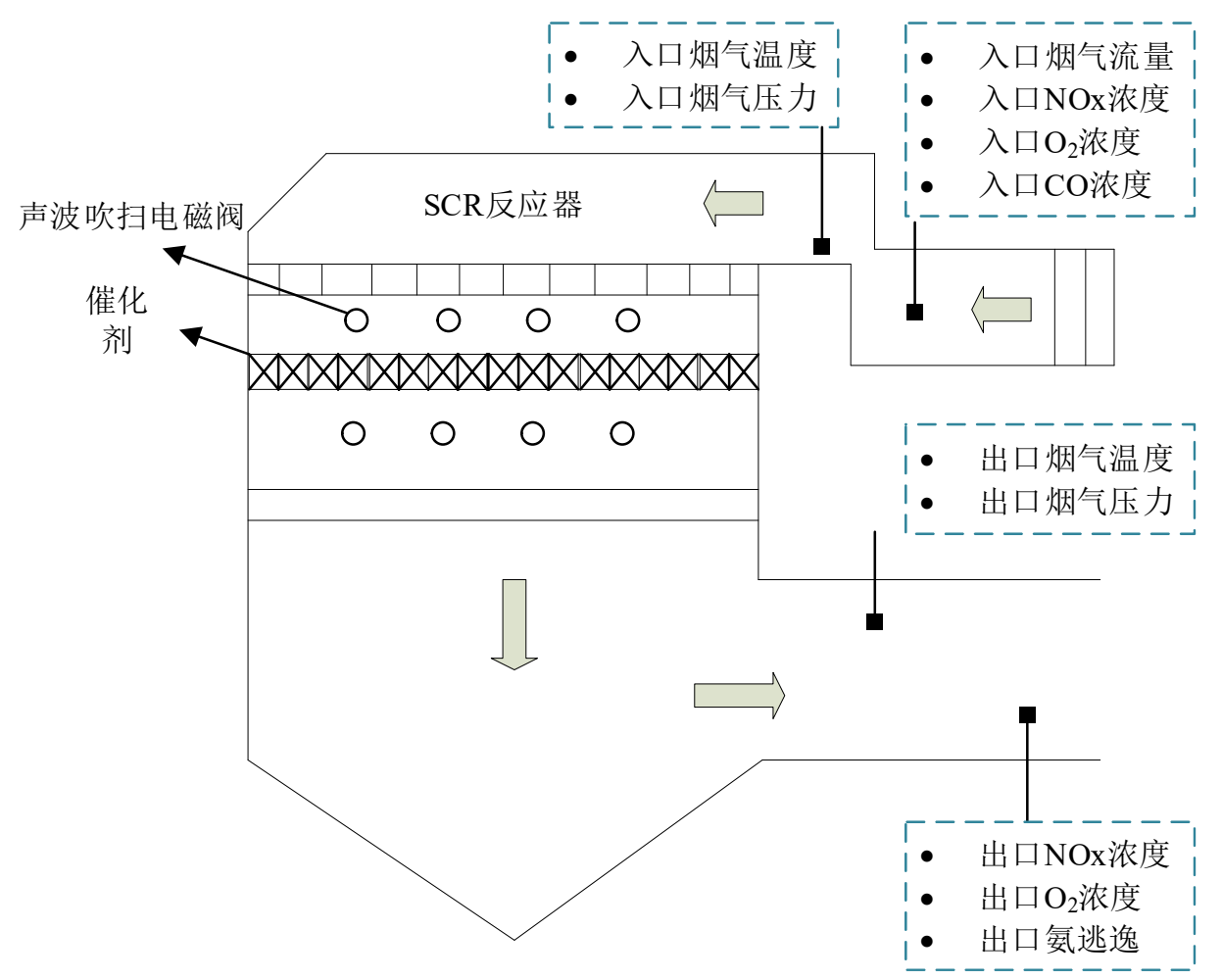

**图 1　电站锅炉 SCR 系统原理图**

**Fig 1　Schematic of the SCR system of a power station boiler**

而减少氮氧化物的排放。吹灰器安装在催化剂上方，其功能是吹散积聚在催化剂上的灰尘，以确保催化剂参与脱硝反应。SCR 反应器可以有效的降低排放烟气中 NOx 的含量，在电厂中得到了广泛的应用[23]。 在 SCR 系统中常用的测点参数也在图 1 中进行展示。

## 2　SCR 动态预测模型

### 2.1　数据预处理

电厂中的分布式控制系统(distributed control system， DCS)和监控信息系统(supervisory information system，SIS)记录了大量历史运行数据，为数据驱动模型的建立提供了基础。将 SCR 出口 NOx 浓度作为预测模型的输出变量，通过机理分析，选择影响输出的 15 个变量作为候选输入变量。以 10s 为间隔共导出 9210 组数据，其中机组负荷处于上下波动的状态。但是由于传感器故障或电磁干扰等因素的影响，这些现场历史数据中含有异常值。采用拉依达准则[24]筛选各变量的异常值，并将该变量异常值出现之前 5 个时刻的平均值代替异常值，以消除异常值对于模型训练的不利影响，处理异常值后的各变量范围如表 1 所示，变量之间的量纲差异会影响模型的训练，因此对数据进行 *min-max* 归一化处理，将变量范围统一到[0,1]，计算公式如下：

$$x' = \frac{x - x_{\min}}{x_{\max} - x_{\min}} \tag{1}$$

其中 $x'$ 是特征量 $x$ 归一化后的数据，$x_{\max}$ 和 $x_{\min}$ 分别是特征量 $x$ 的最大值和最小值。

**表 1　SCR 相关参数的运行范围**

**Table 1　Operating range of the parameters related to SCR**

| 变量名 | 单位 | 变化范围 | 标签 |
|---|---|---|---|
| 出口 NOx 浓度 | mg/Nm3 | 21.378-37.241 | *Y* |
| 入口 NOx 浓度 | mg/Nm3 | 89.921-317.043 | *NOx* |
| 入口 $O_2$ 浓度 | % | 3.673-5.856 | $O2_{in}$ |
| 入口 CO 浓度 | mg/Nm3 | 4.275-864.527 | *CO* |
| 烟气流量 | Nm3/h | 115.333-149.616 | *F* |
| 入口烟气压力 | kPa | -1.155 - -0.549 | $P_{in}$ |
| 入口烟气温度 | ℃ | 350.000-371.427 | $T_{in}$ |
| 加氨量 | m3/h | 43.813-129.658 | *Q* |
| 稀释风机电流 | A | 0.220- 0.708 | *3AB* |
| 出口烟气压力 | kPa | -1.614 - -0.979 | $P_{out}$ |
| 出口烟气温度 | ℃ | 350.876-371.053 | $T_{out}$ |
| 出口 $O_2$ 浓度 | % | 3.566-6.161 | $O2_{out}$ |
| 氨逃逸 | ppm | 1.878-2.379 | *NH3* |
| 机组负荷 | MW | 765.312-967.453 | *Ne* |
| 总煤量 | t/h | 251.343-357.814 | *TF* |
| 总风量 | t/h | 2783.210-3406.900 | *TA* |

### 2.2　相关参数延迟时间

最大信息系数(maximal information coefficient，MIC)是一种衡量变量之间非线性关系的统计指标，且适用于计算大型数据集中变量之间的统计信息。MIC 的计算公式如下：

$$MIC(x,y) = \max_{xy<B(n)} \frac{I(x,y)}{\log_2(\min(x,y))} \tag{2}$$

$$I(x,y) = \sum_{x\in X}\sum_{y\in Y} p(x,y)\log(\frac{p(x,y)}{p(x)p(y)}) \tag{3}$$

其中 $I(x, y)$为变量 $x$ 和 $y$ 之间的互信息值，$p(x)$、$p(y)$为边缘概率分布，$p(x, y)$为联合概率分布，$B(n)$要求设置为数据量 $n$ 的 0.6 次方左右。

SCR 系统出口的 NOx 质量浓度是通过烟气排放连续检测系统(CEMS)测量得到的，由于其中存在较长的伴热管线，导致出口处的 NOx 浓度测量存在一定的滞后时间[25]；同时 SCR 反应器内的脱硝反应及喷氨调节也存在较长延迟时间[26]，因此 SCR 系统具有延迟特性，DCS 数据库中记录的当前 $t$ 时刻的输入变量值并不能准确预测 SCR 系统出口 NOx 浓度 $y(t)$。但当机组正常运行时，由于 CEMS 取样管线长度固定及喷氨调节设备的稳定运行，故因 SCR 出口到烟囱 CEMS 监测点的距离、喷氨调节等因素导致的延迟时间不会发生剧烈变化，通过最大信息系数 MIC 分析建模数据中各变量延迟重构数据与出口 NOx 浓度的非线性相关性，计算各相关参数相对于出口 NOx

**表 2　输入变量和 SCR 出口 NOx 浓度之间的延迟时间**

**Table 2　Delay time between input variables and SCR NOx concentration at the outlet**

| 变量名称 | *NOx* | $O2_{in}$ | *CO* | *F* | $P_{in}$ | $T_{in}$ | *Q* | *3AB* |
|---|---|---|---|---|---|---|---|---|
| 延迟时间 | 170s | 10s | 300s | 50s | 20s | 260s | 440s | 30s |
| 变量名称 | $P_{out}$ | $T_{out}$ | $O2_{out}$ | *NH3* | *Ne* | *TF* | *TA* | — |
| 延迟时间 | 10s | 290s | 10s | 10s | 130s | 110s | 30s | — |

浓度的最终延迟时间，然后进行建模数据集的重构，消除延迟时间对建模精度的影响。

定义建模数据集中的输入变量为 *X*，输出变量为 *Y*，其中 $X=\left[x_1\left(t\right),\cdots,x_m\left(t\right)\right]$，$Y=y\left(t\right)$，$m$ 为输入变量的维数，$t=1,\cdots,N$，$N$ 为建模样本数。首先在总结已有相关文献及锅炉运行专家经验的基础上确定各输入变量 $x_i\left(t\right)$ 的最大延迟时刻 *k*，在延迟时间范围内为变量 $x_i\left(t\right)$ 按照样本采样间隔依次进行延迟重构处理，得到变量 $x_i\left(t\right)$ 的 *k* 组重构数据序列，表示为 $x_i\left(t-1\right),\cdots,x_i\left(t-k\right)$。其次，依次计算 *k* 列数据和输出变量 $y(t)$ 之间的最大信息系数 MIC，对应 MIC 值最大的一组数据序列和输出 $y(t)$ 之间的非线性相关程度最大，其对应的延迟时间即为 $x_i\left(t\right)$ 的最佳延迟时间，选择该列数据作为变量 $x_i\left(t\right)$ 的重构数据序列，变量 $x_i\left(t\right)$ 重构后的维数仍为 1。最终依次求得全部输入变量的延迟时间，得到重构建模数据集。

由于 SCR 系统里的脱硝反应存在较长延迟时间，因此 SCR 系统内的变量较机组其他参数具有更大的延迟范围。设置机组负荷、总风量、总煤量的延迟范围为 300s，其余 SCR 系统内的参数的延迟范围设置为 600s。由于从 DCS 中导出的数据以 10s 为间隔，因此这两类变量的最大延迟时刻分别为 30 和 60。时间延迟的计算结果如表 2，其中 SCR 系统内的喷氨量 *Q* 的延迟时间为 440s，大于总风量 TF 的 110s，验证了计算结果符合现场实际运行情况。

## 2.3　组合特征选择算法

根据脱硝反应的机理分析，SCR 出口 NOx 质量浓度受众多参数的影响。数据驱动模型通常对建模数据敏感，引入冗余信息会降低模型的预测精度和增加训练时间。因此，需要进行特征选择剔除对输出影响较弱的冗余变量。不同特征选择算法由于原理不同导致选出的最佳特征变量集合存在差异，使用单一特征选择算法难以准确全面挖掘影响 SCR 出口 NOx 浓度的最佳特征集合。综合使用多种基于树的特征选择算法设计组合特征选择方案，其中包括分类回归树(classification and regression tree，CART)、随机森林(random forests，RF)、极值梯度提升(extreme gradient boosting，XGBoost)等算法。具体算法步骤如下：

步骤 1：根据机理分析初步确定 15 个影响 SCR 出口 NOx 浓度的变量作为候选特征变量。

步骤 2：分别使用以上三种特征选择算法进行训练，得到 15 个候选变量的重要性排序结果。

步骤 3：将各变量在三种特征选择算法下的重要性结果相加并求均值，作为该变量的最终重要性结果。

步骤 4：根据脱硝反应机理分析保证氮氧化物 $NO_x$、氨气 $NH_3$ 等主要反应物质作为模型的特征变量，然后根据各变量的最终重要性结果，进行多次实验尝试，确定用于后续建模的特征集合。

**表 3　单一特征选择算法的重要性结果(归一化后)**

**Table 3　Sorting results of importance of single features selection algorithm (after normalization)**

| 变量标签 | CART | RF | XGBoost |
|---|---|---|---|
| *NOx* | 0.241 | 0.159 | **0.055** |
| $O2_{in}$ | 0.21 | 0.254 | 0.189 |
| *CO* | 0.305 | 0.096 | 0.068 |
| *F* | 0.176 | 0.032 | 0.019 |
| $P_{in}$ | **0.165** | 0.041 | 0.025 |
| $T_{in}$ | 0.361 | 0.304 | 0.302 |
| *Q* | 1 | 1 | 1 |
| *3AB* | 0 | 0 | 0 |
| $P_{out}$ | 0.186 | 0.021 | 0.029 |
| $T_{out}$ | 0.301 | 0.315 | 0.506 |
| $O2_{out}$ | 0.2 | 0.152 | 0.25 |
| *NH3* | 0.144 | 0.118 | 0.187 |
| *Ne* | 0.249 | 0.151 | 0.218 |
| *TF* | 0.126 | 0.089 | 0.077 |
| *TA* | 0.311 | **0.183** | 0.204 |

表 3 是候选变量在三种特征选择算法下的归一化的重要性结果。从表 3 中可以看出，不同算法对于候选变量的重要性排序不同，例如 CART 算法下出口烟气压力的重要性明显高于其余两种

算法，RF 算法下总风量的重要性明显高于其他两种算法，XGBoost 算法下入口 NOx 浓度的重要性明显低于其他两种算法。这一结果表明单一特征选择算法由于算法原理的差异，在解释某些候选变量重要性时出现较大差异，不足以准确反映各个候选变量对 SCR 出口 NOx 浓度的影响程度。

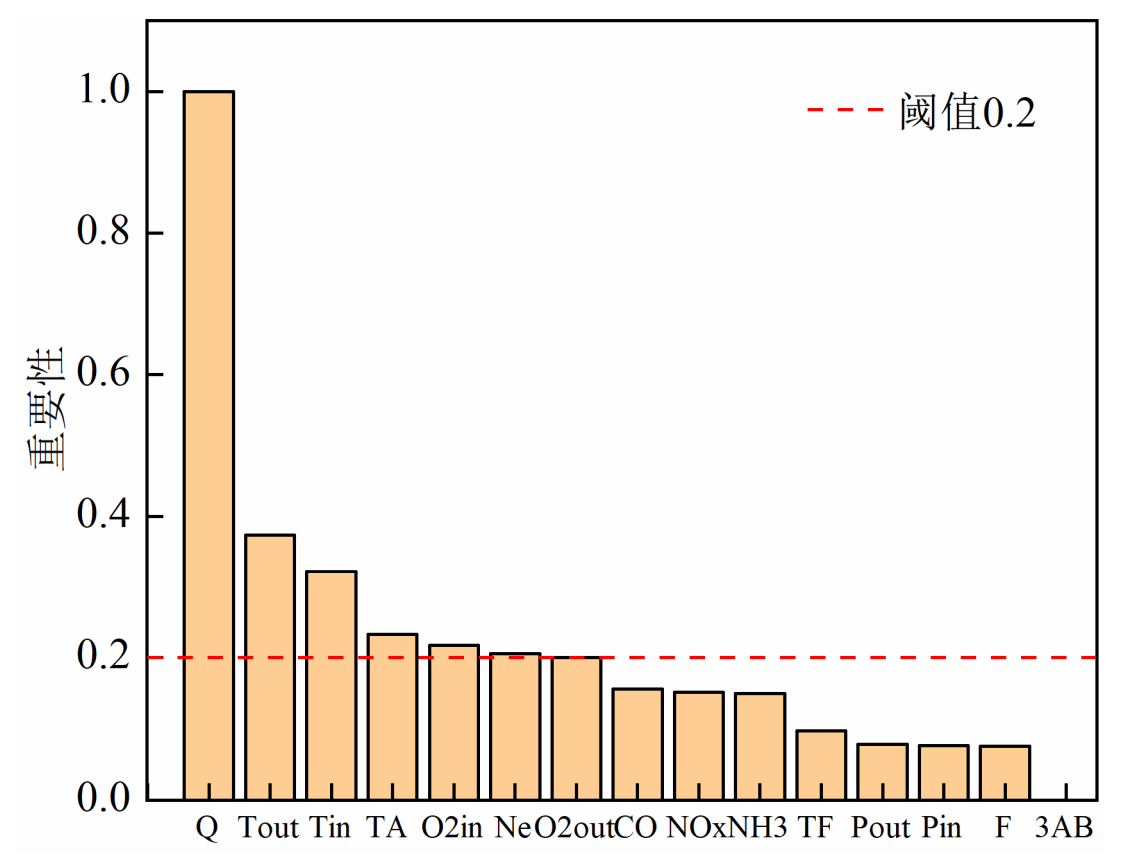

**图 2　组合特征选择算法的重要性排序结果**

**Fig 2　Sorting results of importance by a combined feature selection algorithm**

图 2 是组合特征选择算法(MFS，combined feature selection)的结果，由于 SCR 出口 NOx 是入口 NOx 经脱硝装置处理后产生的，因此保留入口 NOx 浓度作为特征变量；再经实验法确定最终重要性大于 0.2 的候选变量作为模型输入变量，即加氨量 $Q$、出口烟气温度 $T_{out}$、入口烟气温度 $T_{in}$、总风量 $TA$、入口氧气浓度 $O2_{in}$，机组负荷、出口氧气浓度 $O2_{out}$、入口氮氧化物浓度 $NOx$。

## 2.4　变分模态分解

从上述组合特征选择算法的结果得知，加氨量 $Q$ 与 SCR 出口 NOx 浓度的相关程度最高，但是由于 SCR 脱硝反应受到机组负荷、温度、催化剂等众多因素影响，且测量数据中包含传感器测量误差，因此加氨量 Q 具有非平稳性、非线性等特点，其中包含的深层时频信息不明显，增加了神经网络预测模型拟合非线性函数关系的难度。变分模态分解[17] (variational mode decomposition，VMD)是 2014 年提出的一种信号分解方法，可以将信号分解为多个有限带宽的模态分量且可以避免模态重叠[19]。采用 VMD 分解可以将信号分解为指定个数的模态分量 $IMF$，将高频信号视为噪声信号进行剔除，从而达到分离原始信号中不同频率特征的目的。首先初始化待分解信号的模态个数 $K$；然后进行 VMD 分解得到各模态分量 $IMF_1,\cdots,IMF_k$；其次计算最后一个模态分量 $IMF_k$ 和原信号之间的相关系数 $P$；最终当 $P$ 小于提前设定的阈值时，即此时最后一个模态分量 $IMF_k$ 和原信号的相关性较小，不足以表征原信号的时域信息，则停止分解，否则模态个数加 1 继续进行分解，直至满足停止条件。VMD 分解流程图如图 3 所示。将经上述 VMD 分解得到的分量 $IMF_k$ 作为与原始相关性较小的无用分量剔除，使用 $IMF_1,\cdots,IMF_{k-1}$ 代替原信号，从而达到提取原信号深层时域信息的目的。

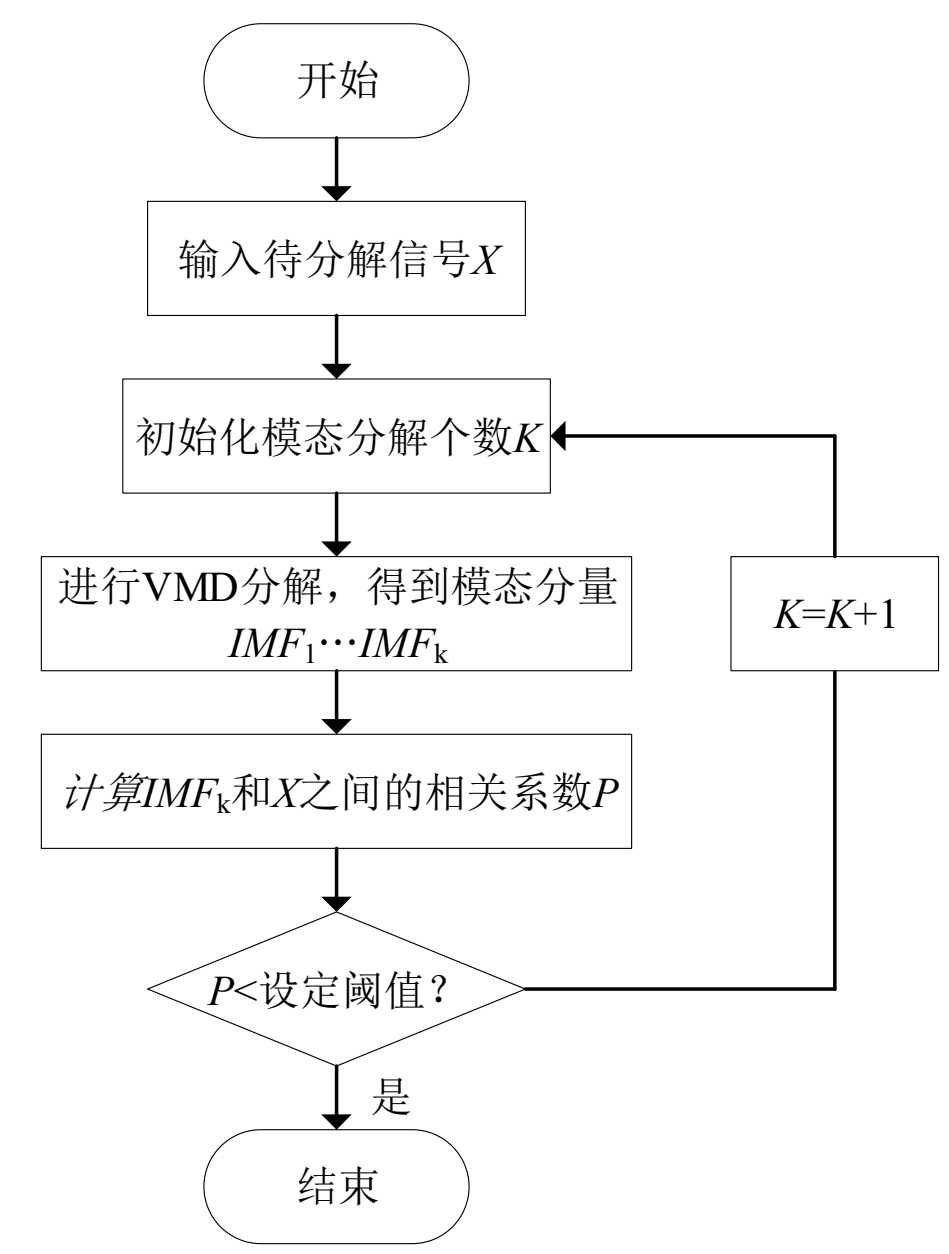

**图 3　VMD 分解流程图**

**Fig 3　Flowchart of VMD**

经过实验，确定加氨量的模态个数为 6，分解结果如图 4 所示。综上，经过组合特征选择算法确定特征变量，并对加氨量 $Q$ 进行 VMD 分解，得到预测模型的输入数据。

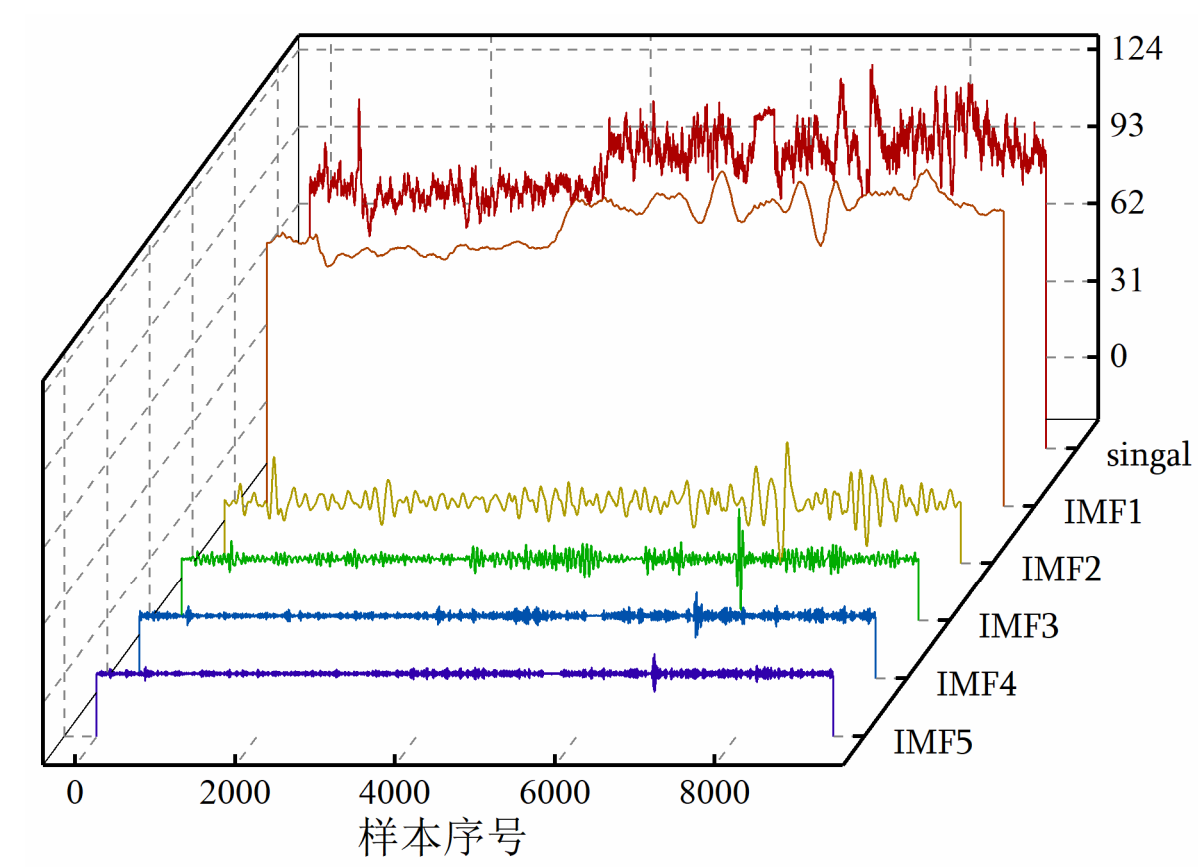

**图 4　加氨量的 VMD 分解结果**

**Fig 4　Decomposition result of ammonia injection volume by VMD**

## 2.5 预测模型

### 2.5.1 极限学习机

极限选学习机[27] (extreme learning machine，ELM)的结构如图 5 所示，是一种单隐层前馈神经网络。假设共有 $N$ 个样本 $(x_i, y_i)$，其中输入样本和输出样本分别为 $x_i=[x_{i1},x_{i2},\cdots,x_{in}]\in R^n$ 、$y_i=[y_{i1},y_{i2},\cdots,y_{im}]\in R^m$，隐藏层神经元的激活函数为 $g(x)$，则 ELM 建模的具体步骤如下[28-29]：

Step1：随机设定输入权重 $w_i$ 和隐藏层偏置 $b_i$，$i=1,\cdots,L$，$L$ 为隐藏层神经元个数。

Step2：计算隐藏层的输出矩阵 $H$。

Step3：计算输出权重 $\beta$：$\beta=H^{\dagger}T$，其中 $H$、$\beta$ 和 $T$ 按照以下公式计算：

$$H\left(w_1,\cdots,w_L,b,\cdots,b_{\tilde{N}},x_1,\cdots,x_N\right)=\begin{bmatrix} g(w_1\cdot x_1+b_1)\cdots g(w_L\cdot x_1+b_L) \\ \vdots \quad\quad \cdots \quad\quad \vdots \\ g(w_1\cdot x_N+b_1) \quad \cdots \quad g(w_L\cdot x_N+b_L) \end{bmatrix}_{N\times L} \tag{4}$$

$$\beta=\begin{bmatrix}\beta_1{}^T \\ \vdots \\ \beta_L{}^T\end{bmatrix} \text{和} \ T=\begin{bmatrix}y_1{}^T \\ \vdots \\ y_N{}^T\end{bmatrix} \tag{5}$$

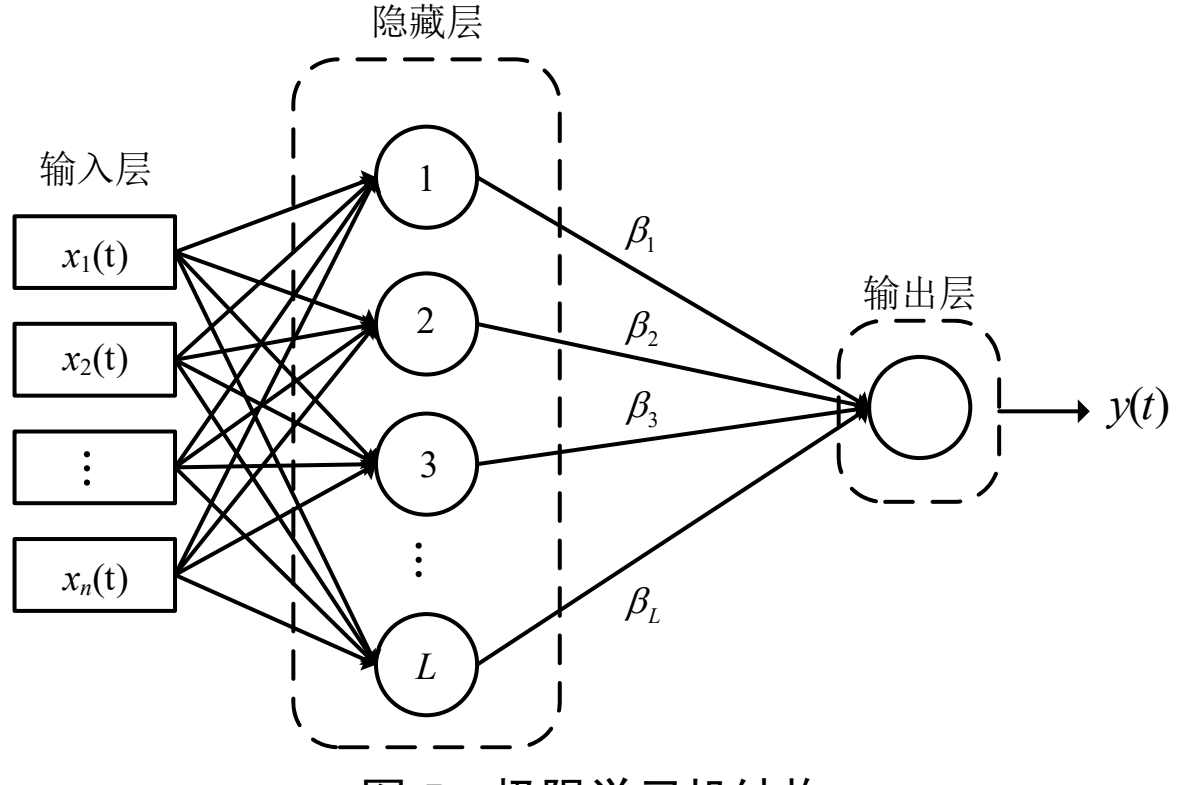

**图 5　极限学习机结构**

**Fig 5　Structure of extreme learning machine**

将原始建模数据经上述组合特征选择算法确定特征变量，并经变分模态分解提取深层时频信息，得到建模数据，建立的预测模型称为 VMD-ELM，则 SCR 出口 NOx 浓度的函数表达式如下：

$$y_p=\sum_{i=1}^{L}[\tanh(w_i\cdot x+b_i)\cdot\beta_i] \tag{6}$$

$$x=[Q_{imf1},Q_{imf2},Q_{imf3},Q_{imf4},Q_{imf5},\\ Tout,Tin,O2in,O2out,TA,Ne,NOx] \tag{7}$$

其中，$y_p$ —SCR 出口 NOx 浓度预测值

$\tanh(x)$ —双曲正切激活函数

$x$ —模型输入数据

$Q_{imf1},\cdots,Q_{imf5}$ —加氨量的模态分量

$w_i\in R^d$ —输入层和隐藏层中第 $i$ 神经元之间的权重

$\beta_i\in R$ —隐藏层中第 $i$ 神经元和输出层之间的权重

### 2.5.2 误差修正模型

为了进一步提升 VMD-ELM 模型的预测精度和使模型适应机组工况的变化，采用误差修正(error correction)策略，即以 VMD-ELM 模型预测误差为研究对象，建立误差修正模型以预测 $t$ 时刻的误差，并叠加 $t$ 时刻 VMD-ELM 模型的预测值，得到最终的 SCR 出口 NOx 浓度的预测值。误差修正模型仍然选用 ELM，以 VMD-ELM 的预测误差为输出变量。预测误差的计算公式如下：

$$e(t)=y_m(t)-y_p(t) \tag{8}$$

其中 $e(t)$ 为 VMD-ELM 的预测误差，$y_p(t)$ 为当前时刻的预测值，$y_m(t)$ 为当前时刻的真实值，并以前 3 时刻的预测误差 $e(t-1)$，$e(t-2)$，$e(t-3)$ 及 VMD-ELM 的原有输入作为误差修正模型的输入。

### 2.5.3 混合预测模型

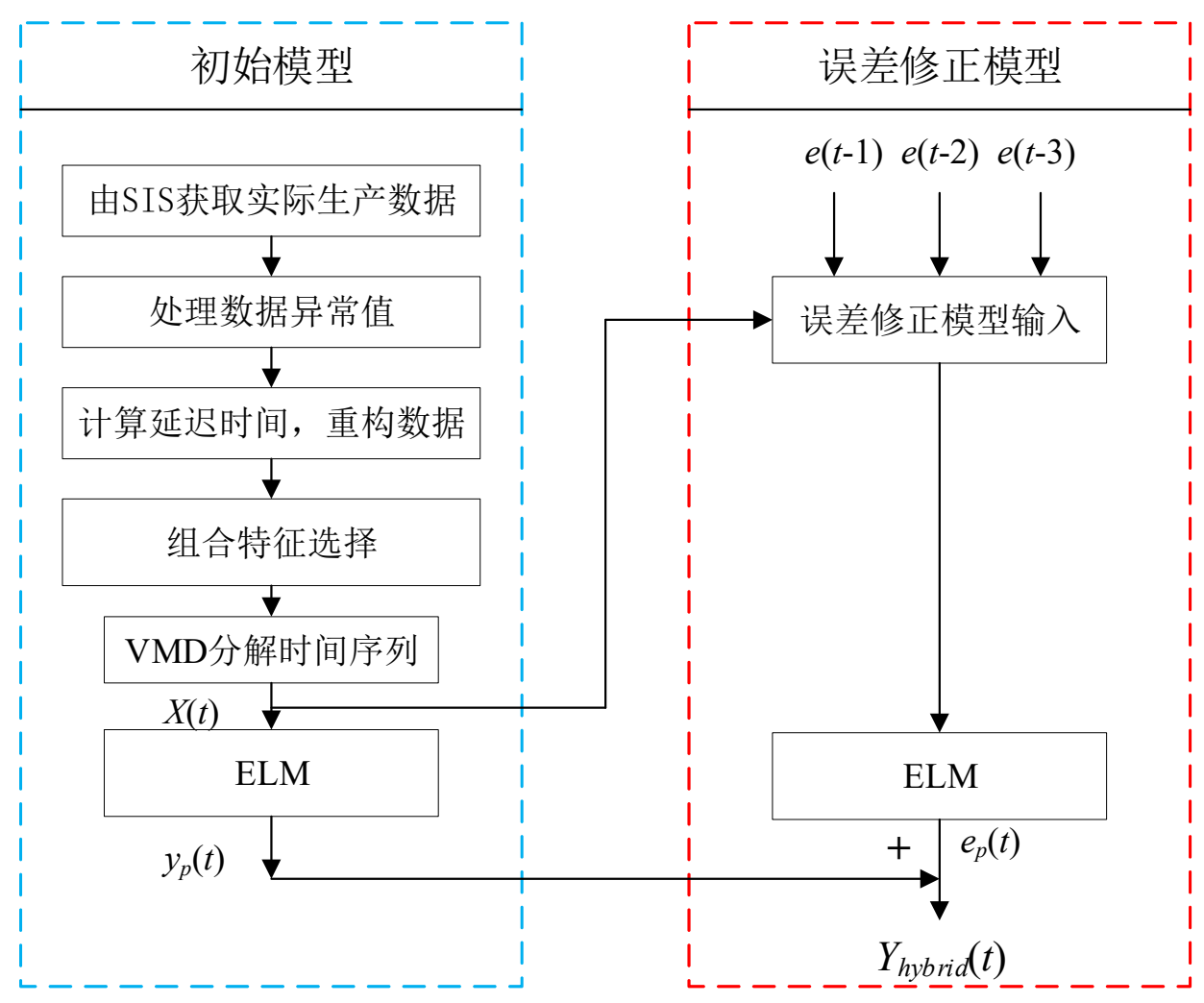

**图 6　SCR 混合预测模型框架图**

**Fig 6　Framework of SCR hybrid predictive model**

将 VMD-ELM 模型和误差修正模型(EC)组成的混合预测模型(EC-VMD-ELM)用于 SCR 出口 NOx 质量浓度预测，其混合建模方案的框架图 6。

建模步骤如下：

(1) 从SIS系统中以10s为间隔导出历史运行数据，并对数据进行预处理。

(2) 利用最大信息系数估算各变量的系统延迟时间，按延迟时间重新对齐数据。

(3) 利用组合特征选取方法确定模型的输入变量。

(4) 对加氨量进行VMD分解，组成模型输入数据$X(t)$。

(5) 利用VMD-ELM模型预测$t$时刻的SCR出口NOx浓度$y_p(t)$。

(6) 将前三时刻的误差$e(t-1)$，$e(t-2)$，$e(t-3)$及输入数据$X(t)$作为输入，利用ELM得到$t$时刻的修正误差$e_p(t)$。

(7) 将$y_p(t)$和修正误差$e_p(t)$叠加得到混合模型的输出值$Y_{hybird}(t)$。

## 3 实验结果与分析

采用数据集的前7500组数据作为训练集，后1710组数据作为测试集。将提出的EC-VMD-ELM模型与DNN[28]、RBFNN[31]、SVR[32]进行对比实验，以验证模型的有效性。在对预测模型进行性能评价时，选择平均绝对误差（MAE）、均方误差（MSE）、平均绝对百分比误差(MAPE)作为评价指标，其计算公式如下[33]：

$$MAE=\frac{1}{N}\sum_{i=1}^{N}\left|Y_i-\hat{Y}_i\right| \tag{9}$$

$$MSE=\frac{1}{N}\sum_{i=1}^{N}\left(Y_i-\hat{Y}_i\right)^2 \tag{10}$$

$$MAPE=\frac{1}{N}\frac{\left|Y-\hat{Y}\right|}{Y}\times 100\% \tag{11}$$

其中$N$为测试集样本个数，$Y_i$为实际测量值，$\hat{Y}_i$为模型预测值。

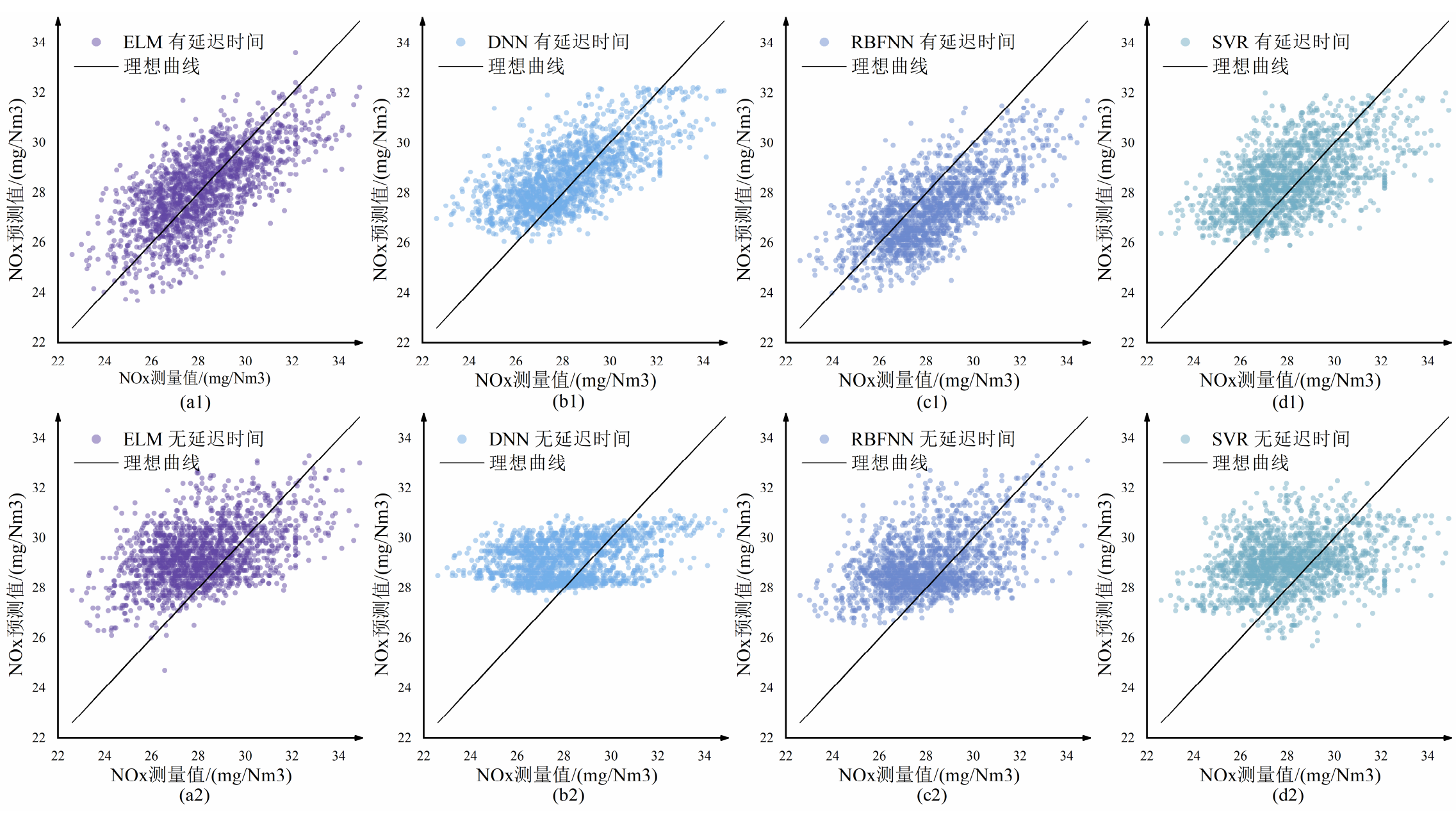

**图7 不同模型的NOx预测值和实际测量值散点图：(a1) (b1) (c1) (d1)考虑延迟时间的数据集，(a2) (b2) (c2) (d2)原始数据集**

**Fig 7 Scatter plot of measured and predicted NOx emissions value of different models : (a1) (b1) (c1) (d1) reconstructed data sets considering time-delay，(a2) (b2) (c2) (d2) original data sets**

**表4 经两种数据集训练的不同模型的NOx排放预测误差对比**

**Table4 Comparisons of NOx emissions predictions errors by different models trained with two data sets**

| 评价指标 | ELM | | DNN | | RBFNN | | SVR | |
|---|---|---|---|---|---|---|---|---|
| | $Y_d$ | $N_d$ | $Y_d$ | $N_d$ | $Y_d$ | $N_d$ | $Y_d$ | $N_d$ |
| MSE | 2.22 | 4.67 | 2.56 | 4.70 | 2.89 | 3.90 | 3.05 | 4.78 |
| MAE | 1.20 | 1.77 | 1.26 | 1.75 | 1.39 | 1.59 | 1.41 | 1.79 |
| MAPE | 4.28% | 6.48% | 4.58% | 6.42% | 4.84% | 5.80% | 5.06 | 6.49% |

### 3.1 相关参数延迟时间对预测结果影响分析

为了验证所提出的计算相关参数延迟时间方法的有效性，本节采用 ELM、DNN、RBFNN、SVR 等算法分别对考虑相关参数延迟时间和不考虑相关参数延迟时间的两种数据集进行建模预测，其中定义考虑延迟时间的数据集为 $Y_{d}$ 和不考虑延迟时间的数据集为 $N_{d}$，数据集均为经过特征选择得到的 8 个输入变量，各个预测模型基于两种数据集建模的参数保持一致。为直观的观察 NOx 预测值和实际测量值的偏差程度，绘制如图 7 所示的测量值和预测值的散点图，图中的黑色线段代表测量值和预测值相等的理想曲线，则测量值和预测值形成的点越靠近理想曲线分布，预测精度越高。由图 7 可知，各算法在未嵌入相关参数延迟时间数据集下的预测值和测量值均存在较大偏差；在考虑延迟时间后，各算法的预测精度均得到改善。

为了定量分析相关参数延迟时间对预测精度的影响，计算两种数据集下的误差评价指标，结果如表 4 所示。根据表中结果得出，4 种算法在数据集 $Y_{d}$ 上的精度较 $N_{d}$ 均得到提升，以 MAPE 为例，ELM 降低 34.0%，DNN 降低 28.7%，RBFNN 降低 16.6%，SVR 降低 22.0%. 其中 ELM 模型精度的提升幅度最大，MSE、MAE、MAPE 均降低 30%以上。通过上述实验结果分析可知，由于原始建模数据集 $N_{d}$ 未考虑延迟时间的影响，相关参数的测量值不能真实反应当前时刻 NOx 排放量，导致建模精度较低；根据相关参数延迟时间重构建模数据集后，提高建模数据之间的非线性相关性，进而提升模型预测精度。

### 3.2 特征选择对预测结果影响分析

为了验证所提出的组合特征选择算法相对于单独使用 CART、RF、XGBoost 算法的有效性，基于组合特征选择算法、CART、RF、XGBoost 等四种算法得到变量重要性大于 0.2 的特征集合，分别建立 ELM 预测模型，图 8 为 ELM 基于不同特征集合得到的 NOx 预测值和实际测量值的散点图。由图 8 可得，基于组合特征选择算法的结果相对于其他算法更贴近理想曲线分布。这主要因为单一特征选择算法的局限性和原理间的差异性，导致其特征集合中漏选对 SCR 出口 NOx 影响较大的变量或增加了冗余变量，例如 RF 特征选择结果中缺少机组负荷、总风量等表征机组锅炉燃烧状态变化的特征变量，该特征的变化将导致锅炉燃烧产生的氮氧化物浓度发生较大波动，进而影响脱硝过程；CART 算法则增加 SCR 入口 CO 浓度作为模型输入变量，因为 SCR 脱硝反应中 CO 不是主要反应物质，故为预测模型引入冗余变量，降低预测精度；XGboost 特征选择结果中缺少入口氧气浓度，但脱硝反应需要氧气及还原剂氨气的参与，故入口氧气浓度影响脱硝反应效率及出口 NOx 浓度。但组合特征选择算法在综合多种特征选择结果的基础上，为预测模型提供了合理的特征集合。

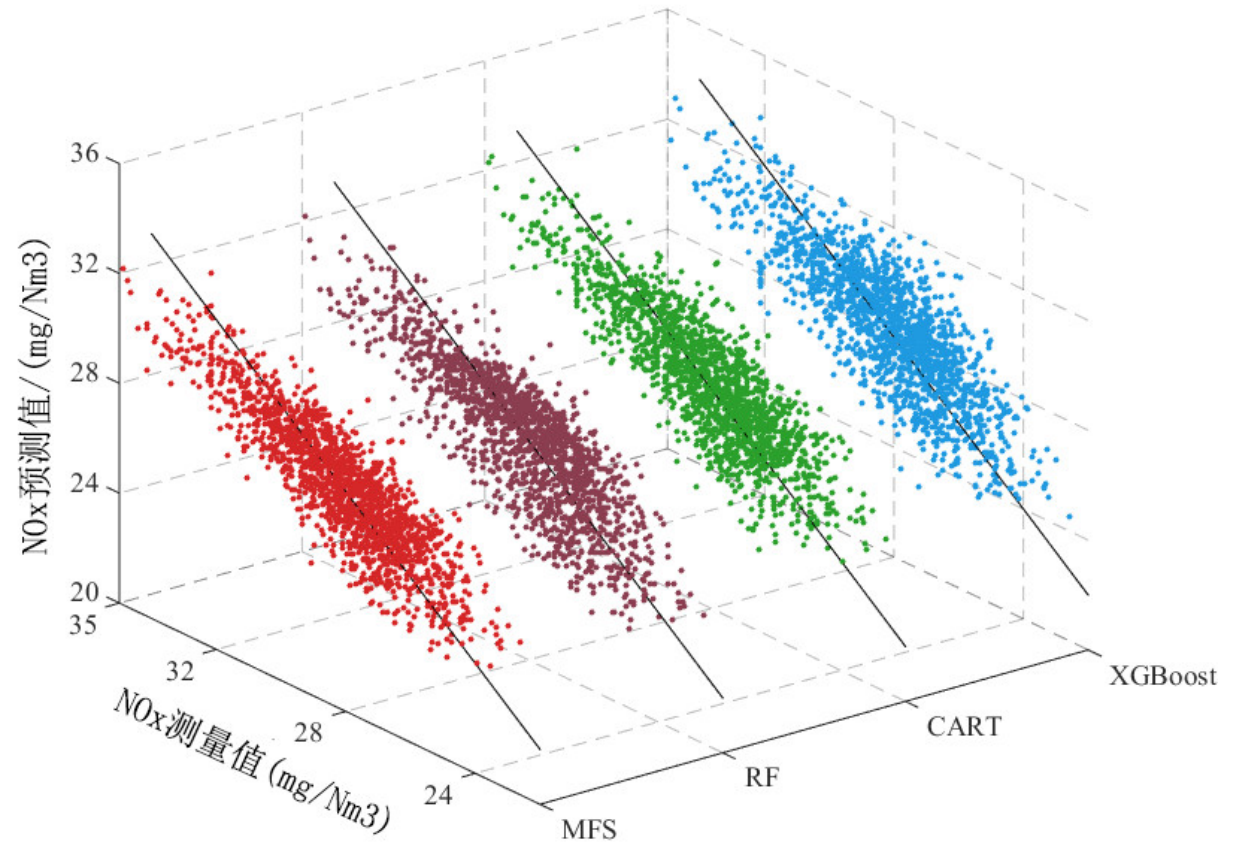

图 8　不同特征选择算法的预测结果

**Fig 8　Prediction results of different feature selection**

表 5　特征选择前后模型误差比较

**Table5　Comparison of model error before and after feature selection**

| 模型 | 方法 | MSE | MAE | MAPE |
|---|---|---|---|---|
| ELM | 特征选择前 | 2.61 | 1.30 | 4.64% |
| | 特征选择后 | 2.22 | 1.20 | 4.28% |
| DNN | 特征选择前 | 2.86 | 1.36 | 4.97% |
| | 特征选择后 | 2.56 | 1.26 | 4.58% |
| RBFNN | 特征选择前 | 3.05 | 1.42 | 5.18% |
| | 特征选择后 | 2.89 | 1.39 | 4.84% |
| SVR | 特征选择前 | 3.38 | 1.50 | 5.42% |
| | 特征选择后 | 3.05 | 1.41 | 5.06% |

为进一步证明组合特征选择算法的适用性，采用 ELM、DNN、RBFNN、SVR 等 4 种算法对是否经过特征选择的数据集进行建模预测，其中特征选择前后的建模数据仅输入变量的维度不同，其余数据预处理及根据相关参数延迟时间重构数据集的处理方法均一致。实验结果的评价指标如表 5 所示。4 种算法在特征选择后的预测结

果更为准确，以ELM为例，其MSE降低了14.6%，MSE 降低了 7.7%，MAPE 降低了 7.8%. 组合特征选择策略为预测模型提供了合理的特征集合，提升各模型的预测精度。

### 3.3 VMD 分解对预测结果影响分析

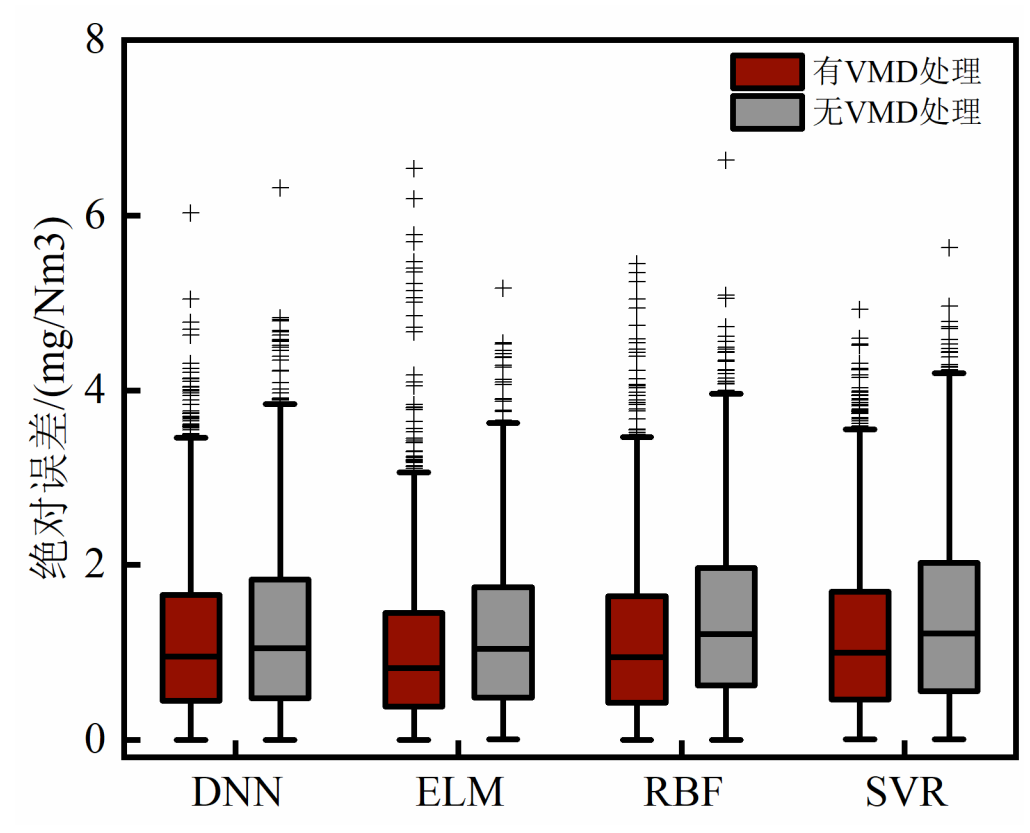

图 9　VMD 分解前后建模绝对误差箱型图

**Fig 9　Box-plot of prediction absolute error of varied models (with and without VMD)**

本节分析 VMD 分解技术对于 4 种建模算法的影响。4 种算法分别基于有 VMD 分解和无 VMD 分解的数据集进行 NOx 排放量预测，其余设置均相同。预测结果如图 9 所示，各模型在 VMD 分解后 NOx 预测值和测量值的绝对误差降低，预测精度提升。氨气作为脱硝反应的还原剂，加氨量直接影响脱硝效率及出口 NOx 浓度，故出现这一实验结果的原因可能是采用变分模态分解技术消除原始加氨信号的非平稳性，挖掘加氨信号中的深层时频信息，增强了非线性预测模型的信息提取能力，有利于预测模型拟合 SCR 出口 NOx 浓度与加氨量及其余变量的深层非线性关系。

### 3.4 EC-VMD-ELM 的预测结果及与其他算法对比

最终建立的 SCR 出口 NOx 预测模型 EC-VMD-ELM 是在 VMD-ELM 的基础上结合误差修正(EC)模型得到的，本节将误差修正前后的结果进行对比，分析误差修正策略对预测结果的影响。表 6 为误差修正前后预测模型的评价指标。经过误差修正后，四种评价指标均有所降低，预测精度得到提升，其中 MAPE 减少 28.1%，MAE 减少 28.4%，MSE 减少 51.4%，出现这一现象的原因是 EC-VMD-ELM 根据误差修正模型的输出对原始预测值进行了误差补充，从而有效缩小预测值和测量值之间的偏差，提高建模精度。

表 6 误差修正前后模型误差比较

**Table6　Comparison of model error before and after error correction**

| 评价指标 | VMD-ELM | EC-VMD-ELM |
| --- | --- | --- |
| MSE | 1.79 | 0.87 |
| MAE | 1.02 | 0.73 |
| MAPE | 3.63% | 2.61% |

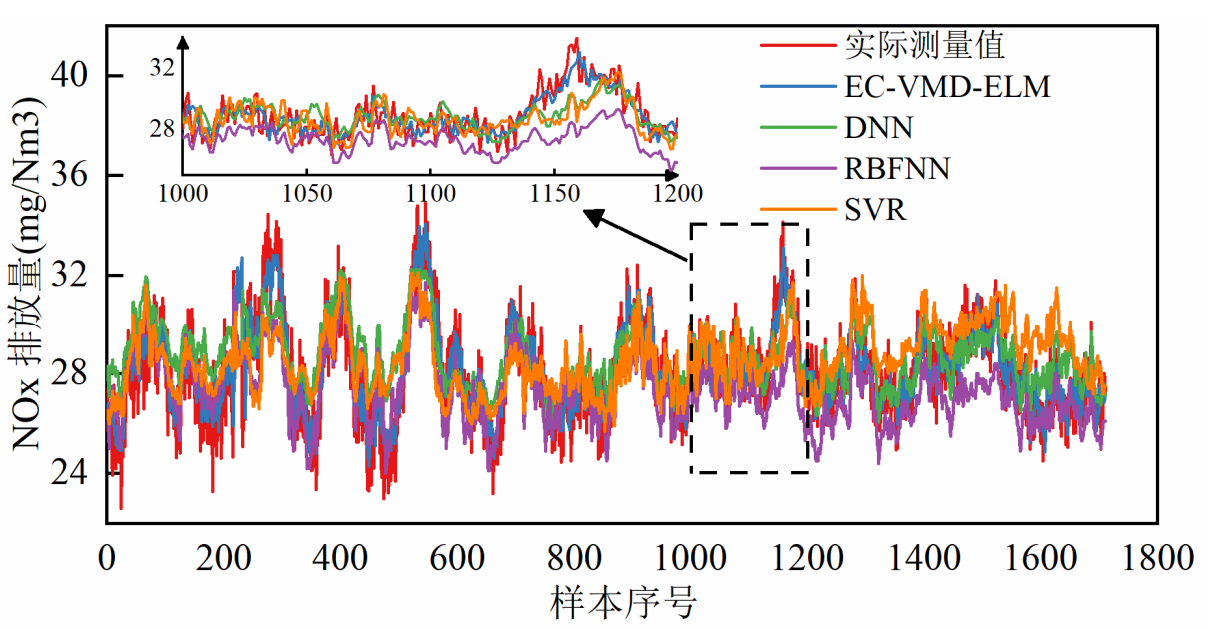

图 10　测试集上的 NOx 排放量预测曲线

**Fig 10　Prediction curves of NOx emission on test set**

通过进行多次实验，先前模型对于测试数据集的平均预测时间为 0.006s，增加修正模型后的平均预测时间为 0.011s，较先前模型的预测时间有所增加，但由于极限学习机的建模预测速度快，修正预测模型仍满足实时在线预测的需要。

为进一步说明 EC-VMD-ELM 算法的准确性，采用经过延迟时间处理和特征选择的变量作为输入，对比 EC-VMD-ELM、DNN、RBFNN、SVR 等模型预测结果。图 10 为 4 种算法预测值和测量值的拟合曲线，其中 EC-VMD-ELM 和对比模型的预测值均符合实际测量值的变化趋势。为进一步对比模型的拟合程度，放大观察其中 200 组数据的拟合曲线。结果显示 EC-VMD-ELM 模型的预测曲线和实际测量曲线之间的偏差最小，拟合效果最优，可以准确预测 SCR 出口 NOx 浓度。而剩余模型的预测值和测量值出现不同程度的偏差，出现这一现象的原因可能是 EC-VMD-ELM 模型在提取深层信息的基础上，对模型预测值进行误差的实时修正，从而提升建模精度；经训练集学习得到的其余对比模型，其模型参数无法适应测试集中出现的新的建模工况数据，由此可见，EC-VMD-ELM 模型相对于其他对比模型具有更好的预测 SCR 出口 NOx 质量浓度的能力。

### 3.5 输入变量的敏感性分析

为进一步研究输入变量的变化对于 SCR 出

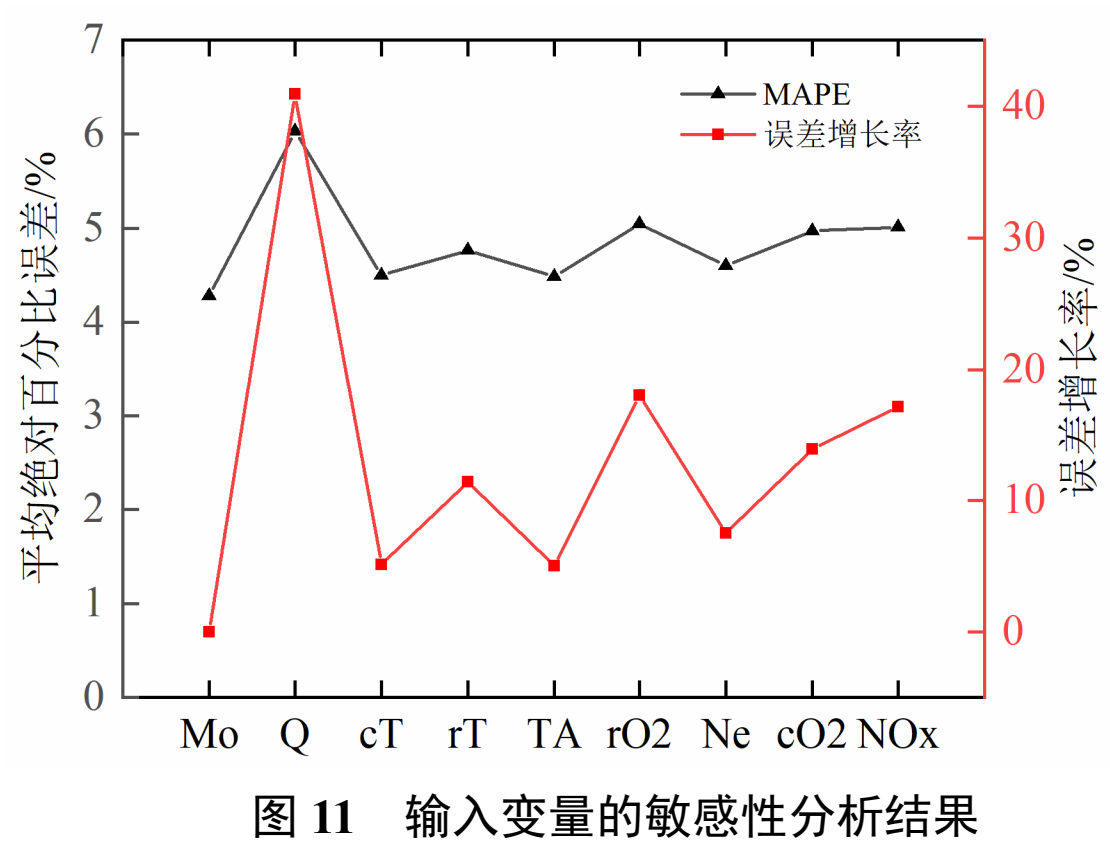

图 11 输入变量的敏感性分析结果

**Fig 11 Sensitivity analysis results of input variables**

口 NOx 浓度的影响，进行输入变量的敏感性分析实验。在建模数据集中各输入变量依次采用均值序列代替的情况下，建立 ELM 预测模型并记录测试集上相应预测结果的 MAPE。在输入变量敏感性分析实验的过程中，训练集和测试集的划分方式、ELM 模型的参数等保持一致，仅研究各输入变量的变化对于输出变量的影响。结果如图 11 所示，横坐标代表各输入变量由均值代替后建立的模型，特别的，由原始数据建立的模型表示为 *Mo*，左侧纵轴代表测试集上预测结果的 MAPE，右侧纵轴代表新建立模型的 MAPE 相对于原始模型 *Mo* 的增长率。根据图 11 可知，当加氨量进行均值代替后，预测误差增加最大，MAPE 相对 *Mo* 增长 40.9%；其次是入口氧气浓度和 SCR 入口 NOx 浓度，MAPE 相对 *Mo* 分别增长 18.0%和 17.1%，入口烟气中的氮氧化物需要结合还原剂氨气及烟气中的氧气进入 SCR 反应器，氮氧化物在催化剂层的作用下被还原为无污染的氮气及水，因此还原剂氨气的使用量影响 SCR 系统的脱硝效果。加氨量过少将导致 SCR 出口氮氧化物的排放量增加，加氨量过多则导则多余的氨气与烟气中的硫氧化物 $SO_3$ 反应生成$(NH_4)_2SO_4$ 等粘性物质，抑制催化剂活性；然后是改变出口氧气浓度和入口烟气温度后，MAPE 相对于原始模型 *Mo* 分别增长 13.9%和 11.4%，这可能由于脱硝化学反应需要氧气及催化剂参与，且烟气温度影响催化剂活性及氨气被氧化的程度，进而影响 SCR 出口 NOx 浓度。在燃煤电厂 SCR 系统喷氨量的调节中，通常考虑出入口 NOx 浓度进行喷氨量设计，进而实现喷氨量值的优化设定。

### 3.6 讨论

通过上述实验分析，得出以下结论：

1）分析相关参数延迟时间可以提升建模精度：SCR 脱硝装置具有大延迟的特点，DCS 系统中相关参数当前时刻的记录值不能真实反映当前时刻 SCR 出口 NOx 浓度。利用 MIC 进行相关参数的时延分析并重构建模数据，增强输入与输出之间的非线性相关性，提升建模精度。

2）组合特征选择算法提升建模精度：单一特征选择算法由于算法原理不同，所选特征集合存在差异且难以全面挖掘对 NOx 排放量影响较大的变量，综合 CART、RF、XGBoost 算法设计组合特征选择算法，为预测模型提供合理的特征集合，减少输入数据中的冗余信息，提高模型预测能力。

3）加氨量分解提升建模精度：通过对加氨量进行 VMD 分解，将加氨量分解为 5 个不同频率的模态分量，提取深层时域信息，使预测模型充分利用其特征信息，提升建模精度。

4）误差修正策略进一步提升建模精度：针对初始模型预测值和测量值的误差，挖掘其变化规律，建立误差预测模型，从而对 NOx 排放量的原始预测值进行修正，较未修正时的 NOx 排放量预测误差降低了 28.1%.

5) 通过对输入变量进行敏感性分析得出，加氨量作为可控变量，其变化幅度对 SCR 出口 NOx 浓度影响显著。此外，也应考虑其他对出口 NOx 浓度影响较大的多种变量，进行智能控制算法的设计。

## 4 结论

针对 SCR 系统变量存在时间延迟和深层时域信息不明显、影响预测模型精度的情况，提出了一种 EC-VMD-ELM 混合建模方法。使用最大信息系数估算各变量的延迟时间，进行数据重构，从而实现 SCR 出口 NOx 浓度的动态预测。并在特征提取和构建预测模型方面进行了研究改进。基于电站历史运行数据进行实验验证，所提出的改进策略均提升了模型的预测精度。EC-VMD-ELM 模型的预测精度优于其余对比算法。在此基础上，通过敏感性分析对 SCR 系统 NOx 排放量影响因素进行分析。下一步工作主要从两个方面开展：一方面，探索其他算法在所提

出框架中应用的有效性和可能性，另一方面，将所提出算法在实际生产中进行推广应用。

## 参考文献

# Dynamic Prediction Model for NOx Emission of SCR System Based on Hybrid Data-driven Algorithms

TANG Zhenhao[1*], WANG Shikui[1], CAO Shengxian[1], Li Yang[2], SHEN Tao[3]

(1. School of Automation Engineering, Northeast Electric Power University; 2. School of Electrical Engineering, Northeast Electric Power University; 3. Harbin Boiler Company Limited)

The NOx emissions of coal-fired power plants cause critical environmental pollution and endanger human health. The SCR system, with the characteristics of easy installation and high denitrification efficiency, has been extensively applied in the post-treatment of the NOx emissions in coal-fired power plants. However, due to nonlinearity, large delay time, and strong disturbance of SCR system, SCR modeling is a challenging problem.

Aiming at predicting NOx emission at the outlet of SCR system, a dynamic prediction model based on hybrid data-driven algorithms is proposed in this paper. The flow of the proposed modeling algorithm is shown in Fig1.

The modeling steps are as follows:

Step1: Collect the practical data from the SIS system. The sample period is 10s.

Setp2: Delete the outliers in the practical data.

Step3: Estimate the delay time between each relative variable and the NOx emissions at the outlet of SCR system. And reconstruct the modeling data according to the delay time.

Step4: The input variables of the prediction model are determined using the combined feature selection method.

Step5: The ammonia injection amount is decomposed by VMD. Combine the decomposition data and the remaining variables to form the input dataset $X$(t).

Step6: Build the VMD-ELM model as initial model to predict the NOx emissions. The initial prediction result is denoted as $y_p(t)$.

Step7: Build the error correction model to correct the prediction result $y_p(t)$ of initial model.

Step8: Get the output of hybrid predictive model by superimposing the initial prediction value and the error correction value.

Experimental results based on practical data show that the *MAPE* of predicted results is 2.61%. The proposed model has high accuracy in predicting the NOx emissions at the outlet of SCR system. In this abstract, Figure 1 is on page 6 of the original paper.

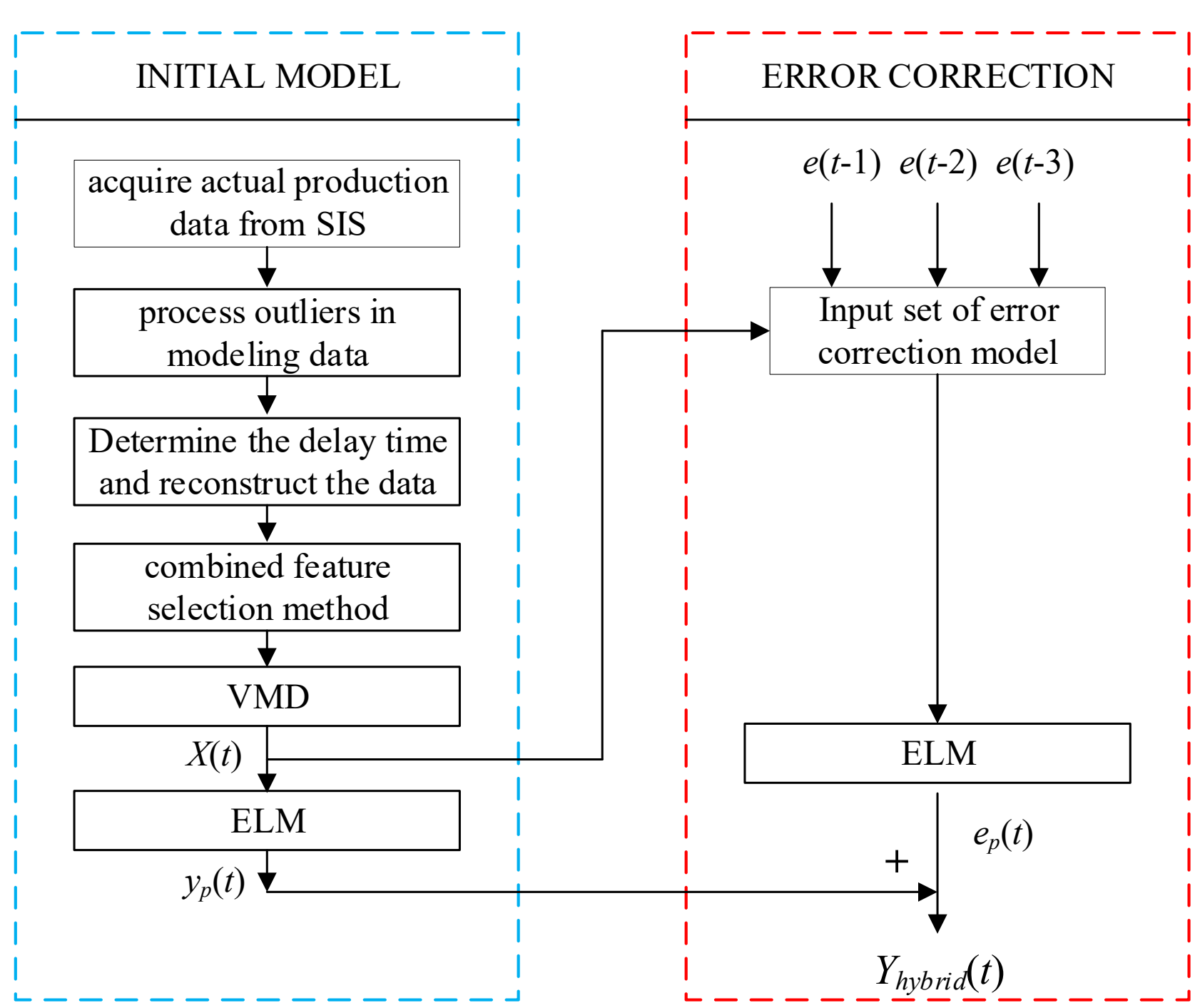

**Fig. 1 Framework of SCR hybrid predictive model**